\documentclass[conference]{IEEEtran}
\IEEEoverridecommandlockouts
\usepackage{cite}
\usepackage{amsmath,amssymb,amsfonts}
\usepackage{algorithmic}
\usepackage{graphicx}
\usepackage{textcomp}
\usepackage{xcolor}
\usepackage{booktabs}
\usepackage{algorithm2e}
\usepackage{inconsolata}
\usepackage{float}
\usepackage[hidelinks]{hyperref}
\def\BibTeX{{\rm B\kern-.05em{\sc i\kern-.025em b}\kern-.08em
    T\kern-.1667em\lower.7ex\hbox{E}\kern-.125emX}}
\begin{document}
\begin{sloppy}

\title{Navigating Complexity in Software Engineering:\\A Prototype for Comparing GPT-n Solutions}

\author{\IEEEauthorblockN{Christoph Treude}
\IEEEauthorblockA{\textit{School of Computing and Information Systems} \\
\textit{The University of Melbourne}\\
Melbourne, Australia \\
christoph.treude@unimelb.edu.au}
}

\maketitle

\begin{abstract}
Navigating the diverse solution spaces of non-trivial software engineering tasks requires a combination of technical knowledge, problem-solving skills, and creativity. With multiple possible solutions available, each with its own set of trade-offs, it is essential for programmers to evaluate the various options and select the one that best suits the specific requirements and constraints of a project. Whether it is choosing from a range of libraries, weighing the pros and cons of different architecture and design solutions, or finding unique ways to fulfill user requirements, the ability to think creatively is crucial for making informed decisions that will result in efficient and effective software. However, the interfaces of current chatbot tools for programmers, such as OpenAI's ChatGPT or GitHub Copilot, are optimized for presenting a single solution, even for complex queries. While other solutions can be requested, they are not displayed by default and are not intuitive to access. In this paper, we present our work-in-progress prototype ``\textsc{GPTCompare}'', which allows programmers to visually compare multiple source code solutions generated by GPT-n models for the same programming-related query by highlighting their similarities and differences.
\end{abstract}


\section{Introduction}

Choosing the right solution to a programming problem is a complex task that depends on a variety of factors, such as project requirements, development team skillset, project deployment environment, and desired level of security, performance, and maintainability. For example, an important consideration is whether to rely on external libraries or APIs~\cite{larios2020selecting}. Although using these can save time and effort in the short term, they can also introduce security vulnerabilities and increase the risk of dependency issues~\cite{zahan2022weak}. Another example is the trade-off between readability and maintainability versus performance~\cite{traini2021software}. Code that is easy to read and maintain is generally preferred, but in real-time systems, for example, high performance is essential and it may be worth sacrificing some readability and maintainability.

When programmers write source code by hand, they are aware of trade-offs and can consider them before writing a new piece of code. However, when artificial intelligence is asked to generate code, the complexity of the trade-offs may be hidden from the programmer, making it difficult to fully understand and evaluate the code that is generated, often with negative consequences. For example, a recent study found that programmers write more insecure code with artificial intelligence assistants, while they are more likely to believe that they wrote secure code~\cite{perry2022users}.

Large language models that power many chatbots for programmers can generate a very large number of solutions for a given task, but current user interfaces may not always provide enough information about these solutions or allow for effective exploration of solution spaces~\cite{treude2022taming}. Instead of informing programmers proactively about the fact that multiple solutions exist and what their pros and cons are, many chatbots simply offer the single solution with the highest probability given the user's input. This can lead programmers to accept the solution presented to them without fully understanding the implications of the code and the differences that other solutions might offer, a phenomenon known as the order effect in recommender systems~\cite{guo2022first}.

In this paper, we argue that it is possible to modify the user interface of a chatbot for programmers in such a way that it shows multiple solutions to a given task and highlights their commonalities and differences. To demonstrate the feasibility of this approach, we introduce an initial prototype of a chatbot for programmers that shows multiple solutions and visually highlights their similarities and differences. This prototype serves as a proof-of-concept and demonstrates that it is possible to create a user interface that provides more information about the different solutions that are generated by artificial intelligence.

\section{Motivating Example}

\begin{figure}
    \centering
    \includegraphics[width=\linewidth]{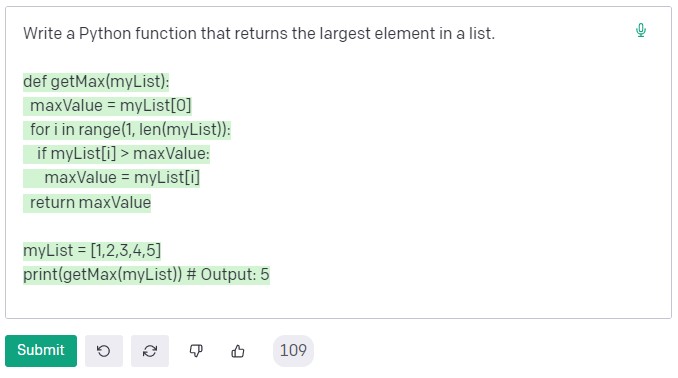}
    \caption{First code snippet generated by GPT-3 for the prompt ``Write a Python function that returns the largest element in a list.''}
    \label{fig:ui1}
\end{figure}

Figure~\ref{fig:ui1} shows an example of the code generated by GPT-3\footnote{\url{https://beta.openai.com/playground}} in response to the prompt ``Write a Python function that returns the largest element in a list.''\footnote{All example tasks used in this paper are taken from \url{https://adriann.github.io/programming_problems.html}.} The code snippet defines a function called \texttt{getMax} that takes a single argument, a list. The function starts by initializing a variable called \texttt{maxValue} to the value of the first element in the list. It then iterates through the remaining elements in the list, comparing each element with the current \texttt{maxValue}. If the current element is larger than \texttt{maxValue}, the \texttt{maxValue} is updated to the value of the current element.

The code snippet also includes an example of how to use the \texttt{getMax} function. It calls the function with an example list containing five elements and provides a source code comment that specifies the expected output of the function. 

It is notable that the code does not make use of any external APIs or libraries, and instead re-implements the functionality for determining the maximum value in a list from scratch. This approach has the advantage of being simple to understand, but may not be as efficient as using a built-in function or library, depending on the size of the input list. Additionally, this approach does not take advantage of code reuse, which could make it less maintainable in the long run.

\begin{figure}
    \centering
    \includegraphics[width=\linewidth]{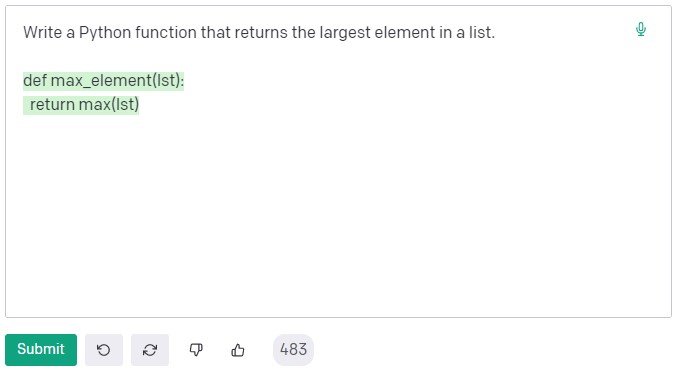}
    \caption{Second code snippet generated by GPT-3 for the prompt ``Write a Python function that returns the largest element in a list.''}
    \label{fig:ui2}
\end{figure}

Upon closer inspection of the GPT-3 interface, it becomes apparent that users can regenerate the response if they are not satisfied with the first response by pressing the button with the two arrows to the right of the \texttt{Submit} button. Pressing this button generates a new code snippet, as shown in Figure~\ref{fig:ui2}.

The code snippet provided in Figure~\ref{fig:ui2} is much shorter than the one provided in Figure~\ref{fig:ui1}. The method name has changed from \texttt{getMax} to \texttt{max\_element}, the name of the argument has changed from \texttt{myList} to \texttt{lst}, but the more significant change is that this code snippet simply uses Python's built-in \texttt{max} function to achieve its task. This approach is more efficient and takes advantage of code reuse, unlike the first snippet. However, unlike the first snippet, it does not provide example input and output.

This example also highlights the limitations of the current GPT-3 interface in terms of navigating multiple solutions. Even with this small programming task, the interface does not provide an easy way to compare different solutions side-by-side, as the old snippet disappears when a new one is generated. The example also demonstrates that even with trivial programming tasks, different ways of solving them exist, in this case with or without reliance on built-in Python functions. 

\begin{figure*}
    \centering
    \includegraphics[width=\linewidth]{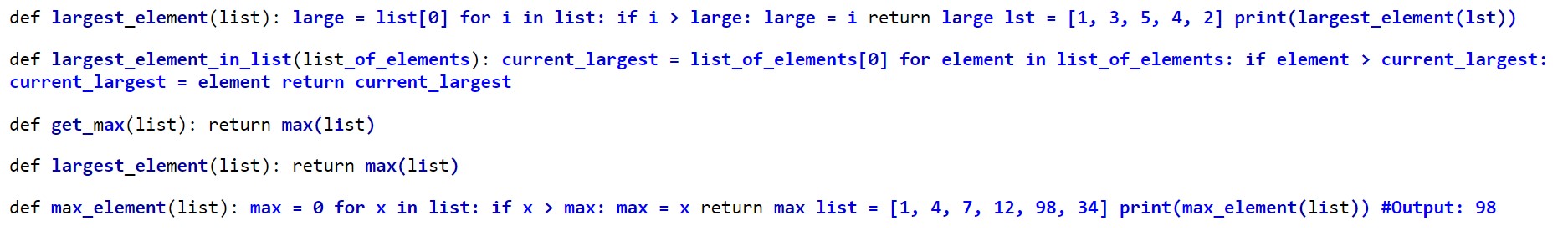}
    \caption{Output generated by \textsc{GPTCompare} for the prompt ``Write a Python function that returns the largest element in a list.''}
    \label{fig:gptcompare}
\end{figure*}

Figure~\ref{fig:gptcompare} illustrates the output of our prototype \textsc{GPTCompare} for the same prompt. Our prototype, by default, displays the first five solutions generated by GPT-3 for a given prompt, but it also adds additional highlighting to indicate the commonalities and differences between the solutions. This allows programmers to easily identify which elements are shared across all solutions and which elements are unique to specific solutions.

In the provided example, all solutions start with the Python keyword \texttt{def}, indicating that they all define a function. Furthermore, all solutions contain an argument whose name starts with \texttt{list}. All solutions contain a \texttt{return} statement, indicating that they all produce a value. These common elements are shown in black.

On the other hand, elements that are not common to all solutions are highlighted with different shades of blue (by default), indicating their uniqueness. For example, the source code comment in the last line (\texttt{Output: 98}) is unique and does not appear in any of the other solutions. It is therefore shown in a bright shade of blue, making it stand out. On the other hand, the call to the built-in \texttt{max} function is less unique in this set of solutions, as both the third and fourth solutions rely on this function to achieve their task. It is shown in a darker shade of blue, indicating that it is not as unique as the first example.

\section{Implementation}

\begin{figure}
    \centering
    \includegraphics[width=\linewidth]{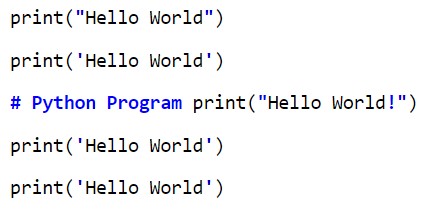}
    \caption{Output generated by \textsc{GPTCompare} for the prompt ``Write a Python program that prints `Hello World' to the screen.''}
    \label{fig:basic}
\end{figure}

Figure~\ref{fig:basic} illustrates the output of \textsc{GPTCompare} in a simple scenario. The prompt given to GPT-3 was ``Write a Python program that prints `Hello World' to the screen.'' GPT-3 generated almost identical code snippets for this prompt, with the exception of one snippet that was preceded by a short source code comment and an exclamation mark added to the end of the string passed to the \texttt{print} function. Upon closer inspection, it is apparent that GPT-3's generated code snippets were not consistent in their use of single quotes or double quotes for strings. Two of the five solutions use double quotes, while the remaining three use single quotes. \textsc{GPTCompare} utilizes color highlighting to distinguish these differences among solutions. The source code comment for the third solution is highlighted in a bright blue color, as is the exclamation mark in the same solution. The different types of quotation marks are highlighted in a darker shade of blue, indicating that they are less unique as they appear in two or three of the solutions, respectively.

\begin{algorithm}[t]
\caption{\textsc{GPTCompare}'s algorithm}
inputs $\gets$ GPT-n.responses\;
n $\gets$ inputs.length\;
\For{i $\gets$ 1, n}{
    diffs $\gets$ []\;
    \For{j $\gets$ 1, n}{
        diff $\gets$ compare(inputs[i], inputs[j])\;
        diffs.append(diff)\;
    }
    m $\gets$ diffs[0].length\;
    \For{k $\gets$ 1, m}{
        uniqueness $\gets$ 0\;
        \ForEach{diff $\in$ diffs}{
            \If{diff[k].startswith(+)}{
                uniqueness += 1\;
            }
        }
        \eIf{uniqueness $>$ 0}{
            output += diff[k]\\(color: 127 + uniqueness * 32)\;
        }
        {
            output += diff[k]\;
        }
    }
}
\end{algorithm}

Algorithm 1 describes the process used by \textsc{GPTCompare} to highlight the differences and similarities between multiple code snippets. It begins by using a diff algorithm such as \texttt{difflib}\footnote{\url{https://docs.python.org/3/library/difflib.html}} to calculate the differences between each pair of input code snippets. In the next phase, the algorithm goes through the output character-by-character and calculates the extent to which each character is unique based on the output of the diff algorithm. Using five code snippets as a starting point, the uniqueness of each character is a number between 0 and 4: If the character also exists in all other solutions, its uniqueness is 0. If it is unique compared to all other solutions (i.e., if it was recognized as an added character in all relevant diffs), its uniqueness is 4. If it was recognized as an added character in only some of the relevant diffs, its uniqueness would be 1, 2, or 3. The color used to display each character is then calculated based on the 256 levels of an RGB color model: if uniqueness is 0, the color will be 0, if uniqueness is larger than 0, the color will be at least 127 (half of the RGB scale) plus a factor multiplied by the uniqueness, for a maximum level equal to the maximum of the RGB scale. 

\section{Related Work}

In this section, we review related work on user interfaces for software bots and on using artificial intelligence to generate source code. Our work lies at the intersection of these research areas.

\subsection{User interfaces for software bots}

A software bot is ``a conduit or an interface between users and services, typically through a conversational user interface''~\cite{storey2016disrupting}. Historically, research on bots for programmers has focused mainly on bots that proactively interact with programmers, such as code review bots integrated into frameworks such as GitHub Actions~\cite{kinsman2021software}. Research has suggested that these bots should aim to provide clear, concise, and well-organized information and focus on appropriate ways of presenting information~\cite{wessel2022guidelines}. Other efforts have focused on the use of bots to help programmers write queries and questions~\cite{liu2022formulate, cao2021automated} and the reduction of the number of steps required to integrate code snippets from online sources into a programmer's code base~\cite{campbell2017nlp2code, reid2021ncq}.

However, less attention has been paid to the user interfaces of such bots for programmers. In our previous work, we have suggested that tools such as GitHub Copilot should prioritize the provision of diverse suggestions rather than redundant ones and explore different methods to highlight commonalities and differences between recommendations~\cite{treude2022taming}. \textsc{GPTCompare} provides a preliminary investigation of these ideas, which are supported by related work from other domains. For example, diversifying reply suggestions for instant messaging systems~\cite{deb2019diversifying} and emails~\cite{buschek2021impact} has been shown to improve creativity~\cite{singh2022hide}.

\subsection{Using artificial intelligence to generate code}

The application of modern natural language processing techniques to programming languages can be traced back to the idea that software is a form of human communication~\cite{sun2022investigating}, known as the naturalness hypothesis~\cite{hindle2016naturalness}. This hypothesis has led to the successful application of natural language processing techniques to source code, such as the adaptation of machine translation techniques to translate source code across different programming languages~\cite{roziere2020unsupervised}, the automated generation of documentation for source code~\cite{feng2020codebert}, or the auto-completion of code~\cite{chen2021evaluating}. These advances are enabled by models such as PLBART~\cite{ahmad2021unified}, CodeBERT~\cite{feng2020codebert}, and Codex~\cite{chen2021evaluating}, the latter of which forms the basis of GitHub's Copilot~\cite{nguyen2022empirical}.

Although these natural language processing techniques have the potential to greatly improve the efficiency and effectiveness of software development, there are concerns regarding their use. One problem with large language models is that they can sometimes ``hallucinate''~\cite{ji2022survey}. A recent study found that a state-of-the-art model was more likely to generate code containing a vulnerability if the query asked for code without that vulnerability~\cite{pearce2022asleep}. Another study found that programmers with artificial intelligence assistants were more likely to believe that they wrote secure code, despite having more insecure code~\cite{perry2022users}. These findings highlight the need for further research on the interface between programmers and the capabilities of large language models, such as GPT-3.

\section{Research Agenda}

\paragraph*{User interface improvements} The initial prototype of \textsc{GPTCompare} has demonstrated the potential to modify the user interface of a chatbot to make it easier for programmers to compare different solutions generated by artificial intelligence. However, further improvements are needed to make it more user-friendly and effective for programmers to use in their daily work. One of the main challenges is to improve the layout of the solutions, as the current prototype does not take into account line breaks. Color settings need to be optimized to help programmers distinguish between unique, common, and essential aspects of the solutions. Another important aspect to consider is the integration of \textsc{GPTCompare} into a code editor, since the current version is a stand-alone prototype.

\paragraph*{Unit of comparison}

An important avenue for future research in tools like \textsc{GPTCompare} is to explore the most effective unit of comparison. In the current prototype, we chose to use character-level comparison, but there are other options that could bring additional benefits. For example, a token-level comparison could ignore simple variable renaming and make the comparison more meaningful. Additionally, depending on the size and complexity of the solutions being compared, line-level or function-level comparisons might be appropriate. Related research on source code representation~\cite{jiang2022hierarchical} has also considered methods such as control flow graphs and program dependence graphs. These approaches could potentially bring even more benefits to the comparison process, but they also present challenges in terms of visual representation and usability for programmers. The most suitable unit of comparison is likely going to depend on programming language and task, as well as user preferences.

\paragraph*{Interactivity}

Future research in code generation should focus not only on creating individual solutions but also on developing interactive tools that allow programmers to seamlessly integrate different aspects of generated code snippets into new, customized solutions. This would eliminate the need for tedious and error-prone manual copy-and-paste, as well as the potential for bugs that can arise from manual integration. The small example in Figures~\ref{fig:ui1} and~\ref{fig:ui2} highlights the potential benefits of such interactivity. Currently, there is no tool support to combine the implementation of the function in Figure~\ref{fig:ui2} with the input and output example in Figure~\ref{fig:ui1}. As the complexity of the code increases, the potential for errors also increases.

\paragraph*{Other software engineering tasks}

Large language models have the potential to not only generate code, but also to assist in debugging by providing explanations for error messages and suggesting fixes. A recent study~\cite{leinonen2022using} found that while Codex-produced explanations were considered ``quite comprehensible'', they were only correct in about half of all cases. Debugging, like code generation, can be a complex task that requires exploring solution spaces. Displaying a single solution or explanation may not be sufficient in many complex scenarios. Therefore, future research should focus on developing tools that assist programmers in effectively navigating the solution space and identifying the most appropriate solutions. 

\paragraph*{Explicit trade-offs}

Providing programmers with the ability to navigate solution spaces is a necessary but not sufficient step towards ensuring that they are able to make informed decisions about the trade-offs associated with different solutions. Future research should also focus on developing tools that provide programmers with additional information that can help them understand and navigate these trade-offs. One approach could be to augment generated solutions with relevant information such as security and performance metrics, possibly in combination with visualizations.

\paragraph*{Reinforcement learning}

It is unlikely that all programmers would have the same preferences for code snippets, even in similar contexts. To address this, chatbots aimed at assisting programmers in exploring solution spaces could use reinforcement learning to learn from the preferences and behavior of their users. This approach would allow the chatbot to adapt to both the overall preferences of the general user population and the individual preferences of specific users.

\paragraph*{Evaluation} 

Evaluating the effectiveness of tools such as \textsc{GPTCompare} requires conducting user studies to understand how programmers interact with existing tools, as well as the approaches proposed in this paper. Different types of users, such as novice programmers and experienced programmers, will likely have different needs and use technology differently depending on their background and level of expertise. User studies can provide valuable insights into the usability and utility of the tools, identify areas for improvement, and tailor the tools to meet the needs of different groups, ultimately helping programmers navigate the complexity of software engineering tasks.


\end{sloppy}

\begin{thebibliography}{10}
\providecommand{\url}[1]{#1}
\csname url@samestyle\endcsname
\providecommand{\newblock}{\relax}
\providecommand{\bibinfo}[2]{#2}
\providecommand{\BIBentrySTDinterwordspacing}{\spaceskip=0pt\relax}
\providecommand{\BIBentryALTinterwordstretchfactor}{4}
\providecommand{\BIBentryALTinterwordspacing}{\spaceskip=\fontdimen2\font plus
\BIBentryALTinterwordstretchfactor\fontdimen3\font minus
  \fontdimen4\font\relax}
\providecommand{\BIBforeignlanguage}[2]{{%
\expandafter\ifx\csname l@#1\endcsname\relax
\typeout{** WARNING: IEEEtran.bst: No hyphenation pattern has been}%
\typeout{** loaded for the language `#1'. Using the pattern for}%
\typeout{** the default language instead.}%
\else
\language=\csname l@#1\endcsname
\fi
#2}}
\providecommand{\BIBdecl}{\relax}
\BIBdecl

\bibitem{larios2020selecting}
E.~Larios~Vargas, M.~Aniche, C.~Treude, M.~Bruntink, and G.~Gousios,
  ``Selecting third-party libraries: The practitioners' perspective,'' in
  \emph{Proceedings of the Joint Meeting on European Software Engineering
  Conference and Symposium on the Foundations of Software Engineering}, 2020,
  pp. 245--256.

\bibitem{zahan2022weak}
N.~Zahan, T.~Zimmermann, P.~Godefroid, B.~Murphy, C.~Maddila, and L.~Williams,
  ``What are weak links in the npm supply chain?'' in \emph{Proceedings of the
  International Conference on Software Engineering: Software Engineering in
  Practice}, 2022, pp. 331--340.

\bibitem{traini2021software}
L.~Traini, D.~Di~Pompeo, M.~Tucci, B.~Lin, S.~Scalabrino, G.~Bavota, M.~Lanza,
  R.~Oliveto, and V.~Cortellessa, ``How software refactoring impacts execution
  time,'' \emph{ACM Transactions on Software Engineering and Methodology},
  vol.~31, no.~2, pp. 1--23, 2021.

\bibitem{perry2022users}
N.~Perry, M.~Srivastava, D.~Kumar, and D.~Boneh, ``Do users write more insecure
  code with {AI} assistants?'' \emph{arXiv preprint arXiv:2211.03622}, 2022.

\bibitem{treude2022taming}
C.~Treude, ``Taming multi-output recommenders for software engineering,'' in
  \emph{Proceedings of the International Conference on Automated Software
  Engineering}, 2022, pp. 1--5.

\bibitem{guo2022first}
X.~Guo, L.~Wang, M.~Zhang, and G.~Chen, ``First things first? {Order} effects
  in online product recommender systems,'' \emph{ACM Transactions on
  Computer-Human Interaction}, 2022.

\bibitem{storey2016disrupting}
M.-A. Storey and A.~Zagalsky, ``Disrupting developer productivity one bot at a
  time,'' in \emph{Proceedings of the International Symposium on Foundations of
  Software Engineering}, 2016, pp. 928--931.

\bibitem{kinsman2021software}
T.~Kinsman, M.~Wessel, M.~A. Gerosa, and C.~Treude, ``How do software
  developers use {GitHub} {Actions} to automate their workflows?'' in
  \emph{Proceedings of the International Conference on Mining Software
  Repositories}, 2021, pp. 420--431.

\bibitem{wessel2022guidelines}
M.~Wessel, A.~Zaidman, M.~A. Gerosa, and I.~Steinmacher, ``Guidelines for
  developing bots for {GitHub},'' \emph{IEEE Software}, 2022.

\bibitem{liu2022formulate}
M.~Liu, X.~Peng, A.~Marcus, C.~Treude, J.~Xie, H.~Xu, and Y.~Yang, ``How to
  formulate specific how-to questions in software development?'' in
  \emph{Proceedings of the Joint European Software Engineering Conference and
  Symposium on the Foundations of Software Engineering}, 2022, pp. 306--318.

\bibitem{cao2021automated}
K.~Cao, C.~Chen, S.~Baltes, C.~Treude, and X.~Chen, ``Automated query
  reformulation for efficient search based on query logs from {Stack}
  {Overflow},'' in \emph{Proceedings of the International Conference on
  Software Engineering}, 2021, pp. 1273--1285.

\bibitem{campbell2017nlp2code}
B.~A. Campbell and C.~Treude, ``{NLP2Code}: Code snippet content assist via
  natural language tasks,'' in \emph{Proceedings of the International
  Conference on Software Maintenance and Evolution}, 2017, pp. 628--632.

\bibitem{reid2021ncq}
B.~Reid, M.~d'Amorim, M.~Wagner, and C.~Treude, ``{NCQ}: Code reuse support for
  {Node.js} developers,'' \emph{arXiv preprint arXiv:2101.00756}, 2021.

\bibitem{deb2019diversifying}
B.~Deb, P.~Bailey, and M.~Shokouhi, ``Diversifying reply suggestions using a
  matching-conditional variational autoencoder,'' in \emph{Proceedings of the
  Conference of the North American Chapter of the Association for Computational
  Linguistics: Human Language Technologies, Volume 2}, 2019, pp. 40--47.

\bibitem{buschek2021impact}
D.~Buschek, M.~Z{\"u}rn, and M.~Eiband, ``The impact of multiple parallel
  phrase suggestions on email input and composition behaviour of native and
  non-native {English} writers,'' in \emph{Proceedings of the Conference on
  Human Factors in Computing Systems}, 2021, pp. 1--13.

\bibitem{singh2022hide}
N.~Singh, G.~Bernal, D.~Savchenko, and E.~L. Glassman, ``Where to hide a stolen
  elephant: Leaps in creative writing with multimodal machine intelligence,''
  \emph{ACM Transactions on Computer-Human Interaction}, 2022.

\bibitem{sun2022investigating}
J.~Sun, Q.~V. Liao, M.~Muller, M.~Agarwal, S.~Houde, K.~Talamadupula, and J.~D.
  Weisz, ``Investigating explainability of generative {AI} for code through
  scenario-based design,'' in \emph{Proceedings of the International Conference
  on Intelligent User Interfaces}, 2022, pp. 212--228.

\bibitem{hindle2016naturalness}
A.~Hindle, E.~T. Barr, M.~Gabel, Z.~Su, and P.~Devanbu, ``On the naturalness of
  software,'' \emph{Communications of the ACM}, vol.~59, no.~5, pp. 122--131,
  2016.

\bibitem{roziere2020unsupervised}
B.~Roziere, M.-A. Lachaux, L.~Chanussot, and G.~Lample, ``Unsupervised
  translation of programming languages,'' \emph{Advances in Neural Information
  Processing Systems}, vol.~33, pp. 20\,601--20\,611, 2020.

\bibitem{feng2020codebert}
Z.~Feng, D.~Guo, D.~Tang, N.~Duan, X.~Feng, M.~Gong, L.~Shou, B.~Qin, T.~Liu,
  D.~Jiang \emph{et~al.}, ``Codebert: A pre-trained model for programming and
  natural languages,'' \emph{arXiv preprint arXiv:2002.08155}, 2020.

\bibitem{chen2021evaluating}
M.~Chen, J.~Tworek, H.~Jun, Q.~Yuan, H.~P. d.~O. Pinto, J.~Kaplan, H.~Edwards,
  Y.~Burda, N.~Joseph, G.~Brockman \emph{et~al.}, ``Evaluating large language
  models trained on code,'' \emph{arXiv preprint arXiv:2107.03374}, 2021.

\bibitem{ahmad2021unified}
W.~U. Ahmad, S.~Chakraborty, B.~Ray, and K.-W. Chang, ``Unified pre-training
  for program understanding and generation,'' \emph{arXiv preprint
  arXiv:2103.06333}, 2021.

\bibitem{nguyen2022empirical}
N.~Nguyen and S.~Nadi, ``An empirical evaluation of {GitHub} {Copilot's} code
  suggestions,'' in \emph{Proceedings of the International Conference on Mining
  Software Repositories}, 2022, pp. 1--5.

\bibitem{ji2022survey}
Z.~Ji, N.~Lee, R.~Frieske, T.~Yu, D.~Su, Y.~Xu, E.~Ishii, Y.~Bang, A.~Madotto,
  and P.~Fung, ``Survey of hallucination in natural language generation,''
  \emph{ACM Computing Surveys}, 2022.

\bibitem{pearce2022asleep}
H.~Pearce, B.~Ahmad, B.~Tan, B.~Dolan-Gavitt, and R.~Karri, ``Asleep at the
  keyboard? {A}ssessing the security of {GitHub} {Copilot's} code
  contributions,'' in \emph{Proceedings of the Symposium on Security and
  Privacy}, 2022, pp. 754--768.

\bibitem{jiang2022hierarchical}
Y.~Jiang, X.~Su, C.~Treude, and T.~Wang, ``Hierarchical semantic-aware neural
  code representation,'' \emph{Journal of Systems and Software}, vol. 191, p.
  111355, 2022.

\bibitem{leinonen2022using}
J.~Leinonen, A.~Hellas, S.~Sarsa, B.~Reeves, P.~Denny, J.~Prather, and B.~A.
  Becker, ``Using large language models to enhance programming error
  messages,'' \emph{arXiv preprint arXiv:2210.11630}, 2022.

\end{thebibliography}
\end{document}